\documentclass[fleqn,twoside]{article}
\usepackage{espcrc2}
\usepackage{amssymb}
\usepackage{latexsym}
\usepackage[dvips]{graphicx}
\usepackage{psfrag}

\title{Topology and Staggered Fermion Action Improvement}

\author{Kit Yan Wong
        \address[MCSD]{Physics Department, Simon Fraser University,
	               Burnaby, B.C., Canada, V5A 1S6} 
        and
        R. M. Woloshyn
        \address[MCSD]{TRIUMF, 4004 Westbrook Mall,
                       Vancouver, B.C., Canada, V6T 2A3}}
       
\begin{document}

\begin{abstract}
It is conventional wisdom that staggered fermions do not
feel gauge field topology. However, the response of staggered
fermion eigenmodes to the topology of the gauge field can
depend quite sensitively on the way in which the staggered
fermion action is improved. We study this issue using a
variety of improved staggered quark actions. We observe
that the separation between the ``would be'' zero modes
and the non-chiral modes increases with the level of improvement.
This enables the ``zero modes'' to be identified
unambiguously. The distribution of the remaining non-chiral modes
is compared with the predictions of Random Matrix Theory.
Satisfactory agreement is obtained.
\vspace{1pc}
\end{abstract}

\maketitle

It is a long standing problem that staggered fermions do not
feel gauge field topology on coarse lattices.
In particular, studies of the microscopic
eigenvalue spectrum of the staggered Dirac operator, upon
comparison with the analytic results predicted by Random
Matrix Theory (RMT), show that only the trivial topological
charge sector is probed by the operator \cite{Damgaard2000}.
On the other hand, it has been suggested that lattice artifacts,
in particular flavour changing interactions associated with
the staggered quark action, are the main cause for this failure
and other related topological problems \cite{Kogut1998}.
The staggered quark action describes four quark flavours in
the continuum limit and the eigenvalue spectrum has a
4-fold degeneracy in this limit. At finite lattice spacing,
the flavour changing interactions break the flavour symmetry
and the degeneracy is lifted. Consequently, staggered fermions
do not have exact zero modes at finite lattice spacing because
the continuum chiral modes (if there are any) are scattered on
the lattice. One thus expects staggered fermions to show
better topological properties if flavour changing effects can 
be suppressed.

In this project, this issue is examined using a variety of
improved staggered quark operators, which are designed to
suppress flavour changing effects by smoothing out
the quark-gluon interaction vertex \cite{Lepage}.
We observe that the ``would be'' zero modes are visibly
decoupled from the non-chiral modes as the level of
improvement increases. The effect of changing the lattice
spacing on the eigenvalue spectrum is also studied.
We again observe that separation between the ``zero modes''
and the non-chiral modes increases as one approaches the
continuum limit. This enables the ``zero modes'' and
subsequently the gauge field topological indices to be
identified. Although the charge indices obtained by using
different operators do not agree on a configuration
by configuration basis, the charge distributions are
found to have no significant difference. Finally,
distribution of the remaining non-chiral modes
is compared with the predictions of RMT.

The unimproved staggered Dirac operator is
\begin{equation}
D_{x,y} = \frac{1}{2} \sum_{\mu} \eta_{\mu}(x) 
          \left[ U_{\mu}(x)\delta_{x+\mu,y}-
                 U_{\mu}^{\dagger}(y)\delta_{x-\mu,y} \right],
\label{unimpaction}
\end{equation}
where $\eta_{\mu}(x)=(-1)^{x_{1}+\ldots+x_{\mu-1}}$ is the
standard fermion phase. A variety of improvement schemes are
considered here. The basic improved version
is the $\mathcal{O}(a^{2})$ improved Asq operator
\cite{Orginos1999}, which includes an additional 3-link Naik
term and replaces the gauge field in (\ref{unimpaction}) with
Fat7 effective links (sum of the original link and the
nearby paths, up to 7-link staples). Further improvement
iterates this fattening procedure with an
additional SU(3) unitarization step between successive
iterations, giving the improved Asq operators
$(\mathrm{UFat7})^{n}\times\mathrm{Asq}$ \cite{Follana2003}.
The HYP-improved operators $(\mathrm{HYP})^{n}$
\cite{Hasenfratz2001} are constructed in the similar fashion
but only those links within the hypercube containing the
original link are included in the fattening process.
We refer the readers to the respective papers for
further details.

Simulations are done with the standard Wilson gauge field action at
three values of coupling, $\beta=5.85, 6.0$ and $6.2$.
About $1000$ configurations are generated for each $\beta$.
The lattice sizes are $10^{4}$, $12^{4}$ and $16^{4}$
respectively so that the physical volumes are
$\sim(1.2fm)^{4}$ in all cases. We compute the eigenvalues
($\lambda^{2}$) and chirality ($\chi$) of the lowest $10$
eigenstates of $-D^{2}$ for all operators listed above.
The corresponding eigenvalues of $D$ are $\pm i\lambda$.
Since $D^{2}$ connects either even-even or odd-odd sites
on the lattice, only half of the spectrum is computed.
For comparison, the lowest $5$ eigenvalues
(in each chiral sector) are also computed for the overlap
operator \cite{Neuberger1998} on the $10^4$ lattice.
\begin{figure}
\begin{minipage}[b]{8.5pc}
\psfrag{gname}[rb][rb]{unimproved}
\psfrag{xaxis}[ct][ct]{\tiny $\lambda$}
\psfrag{yaxis}[cb][cb]{\tiny $|\chi|$}
\includegraphics[width=5.9pc,angle=-90]{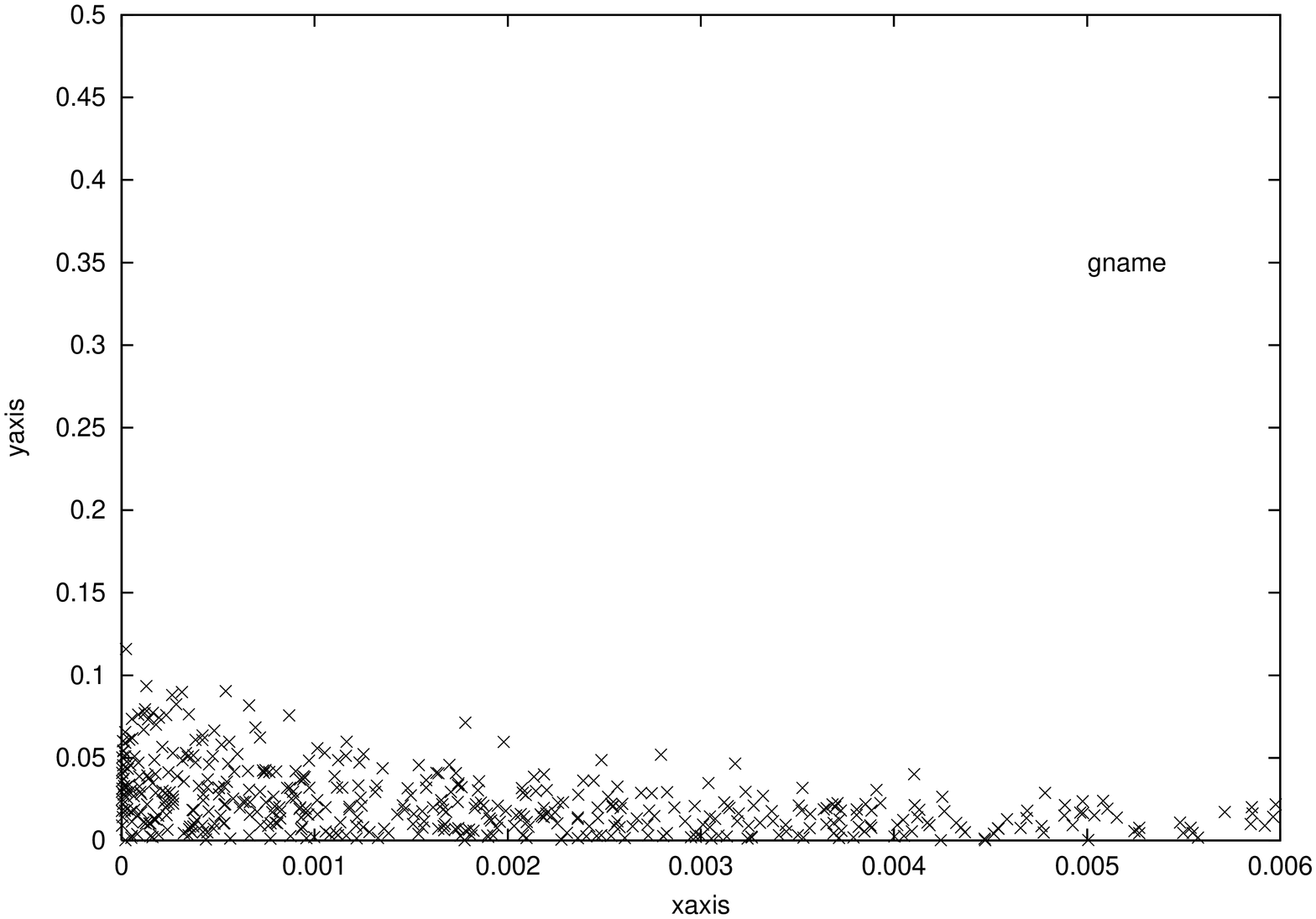}
\end{minipage}%
\begin{minipage}[b]{8.5pc}
\psfrag{gname}[rb][rb]{Asq}
\psfrag{xaxis}[ct][ct]{\tiny $\lambda$}
\psfrag{yaxis}[cb][cb]{\tiny $|\chi|$}
\includegraphics[width=5.9pc,angle=-90]{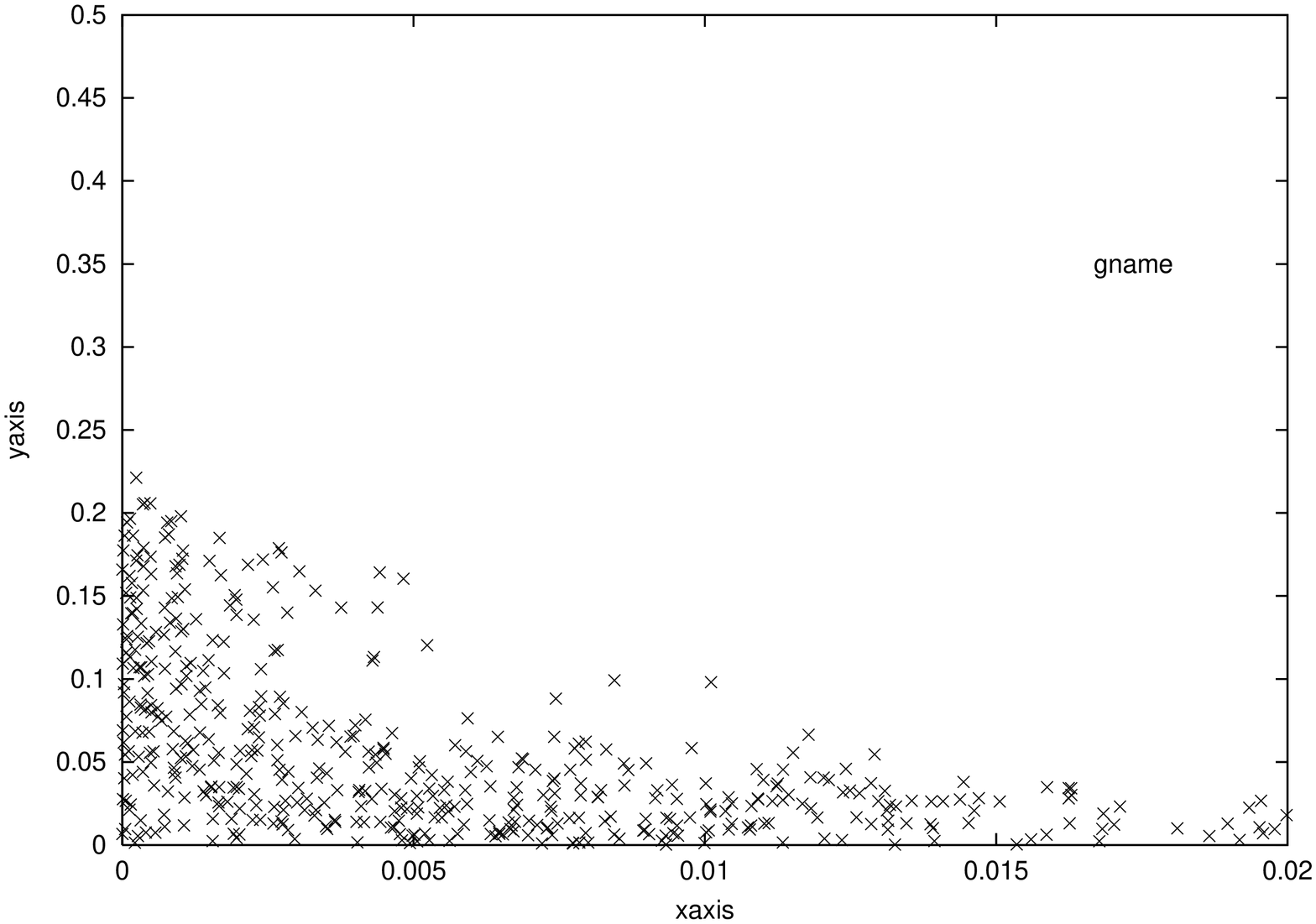}
\end{minipage}\\[0.5pc]
\begin{minipage}[b]{8.5pc}
\psfrag{gname}[rb][rb]{$\mathrm{UFat7}\times\mathrm{Asq}$}
\psfrag{beta}[rb][rb]{$\beta=5.85$}
\psfrag{xaxis}[ct][ct]{\tiny $\lambda$}
\psfrag{yaxis}[cb][cb]{\tiny $|\chi|$}
\includegraphics[width=5.9pc,angle=-90]{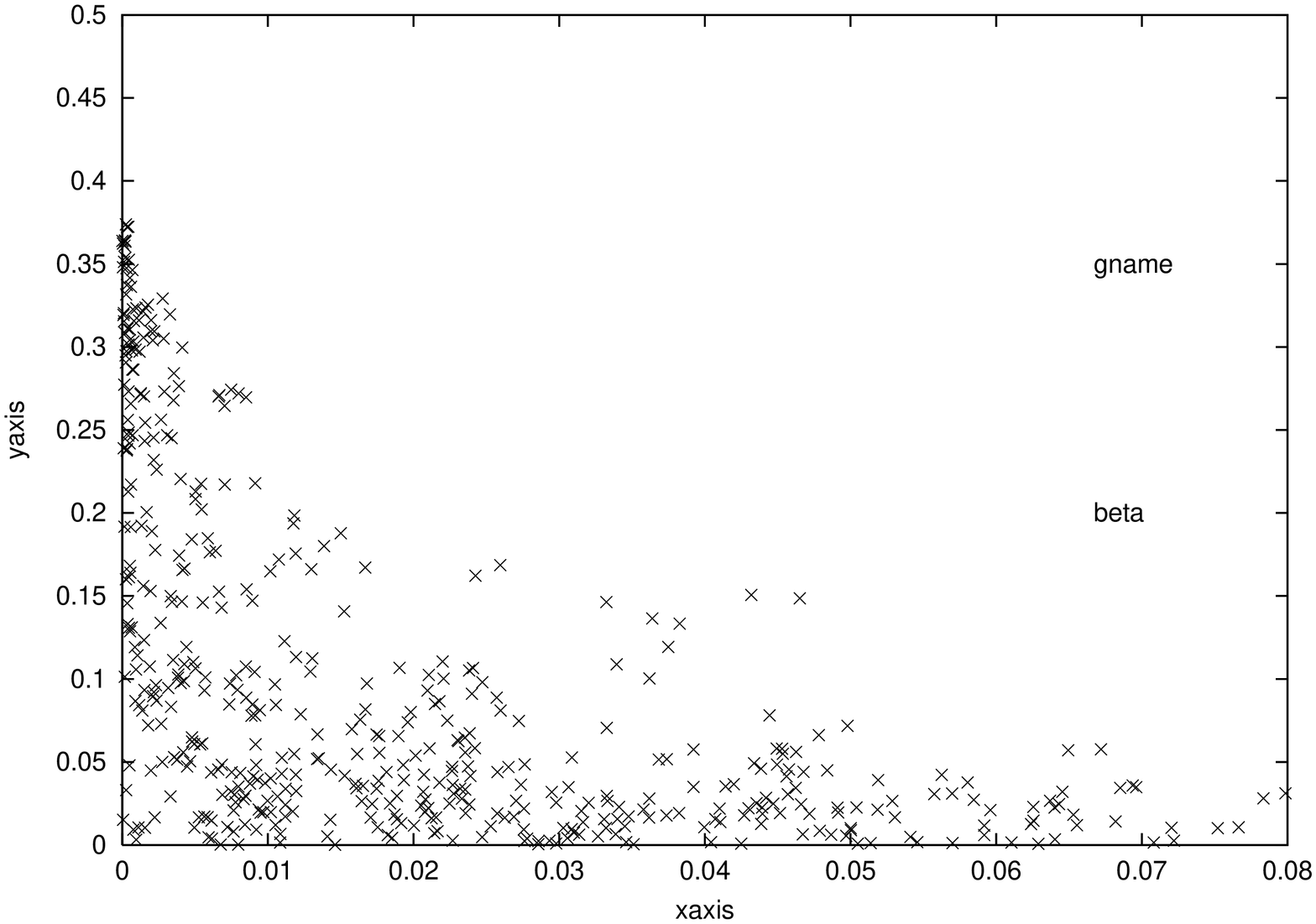}
\end{minipage}%
\begin{minipage}[b]{8.5pc}
\psfrag{gname}[rb][rb]{$\mathrm{HYP}$}
\psfrag{xaxis}[ct][ct]{\tiny $\lambda$}
\psfrag{yaxis}[cb][cb]{\tiny $|\chi|$}
\includegraphics[width=5.9pc,angle=-90]{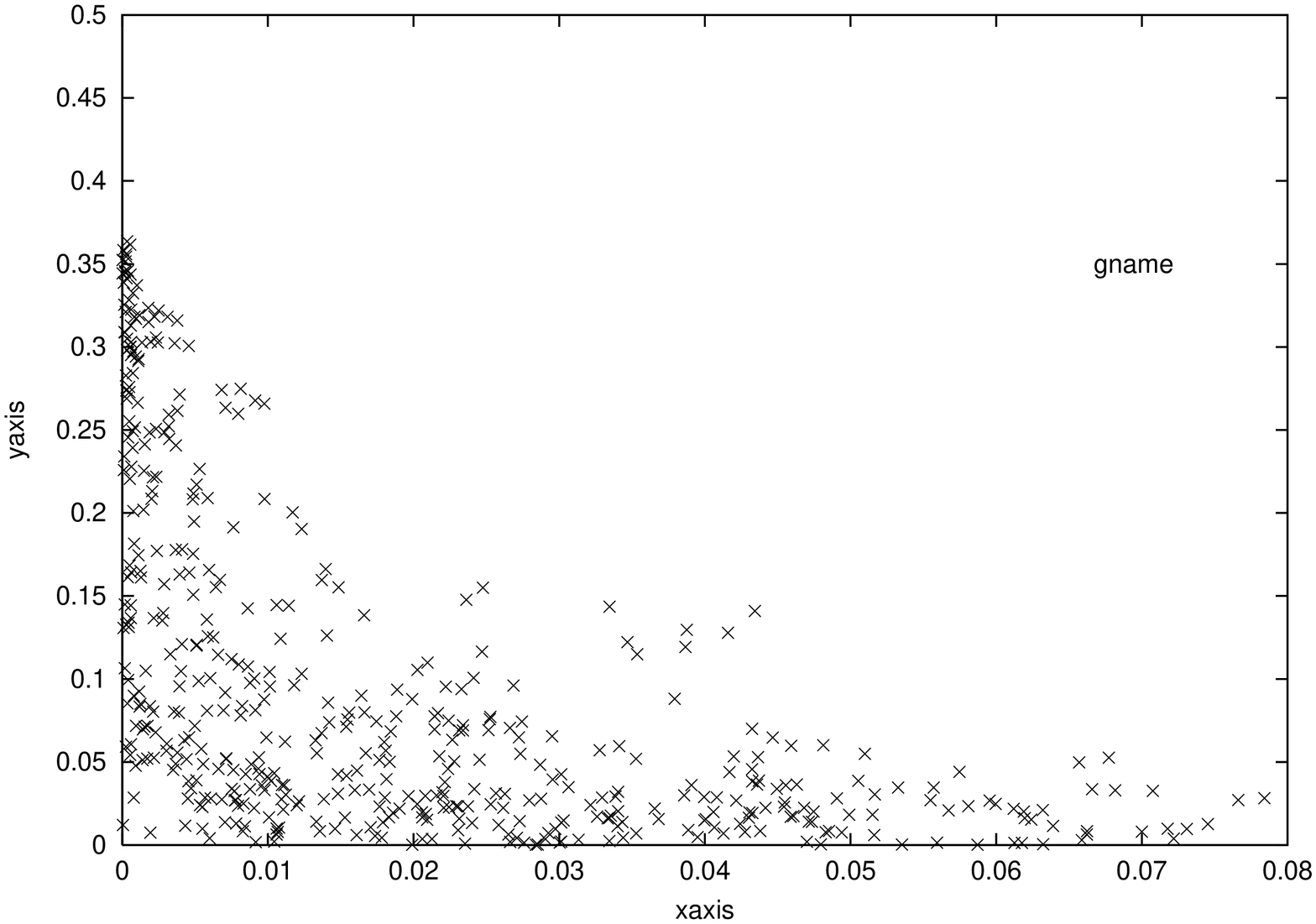}
\end{minipage}
\caption{Eigenvalue spectra of different operators
at $\beta=5.85$. Results are shown for $50$ configurations.}
\label{spectrum}
\end{figure}

The effect of improvement on the eigenvalue spectrum
is shown in figure \ref{spectrum} where $|\chi|$ is plotted
against $\lambda$ for the different operators at a fixed
coupling $\beta=5.85$.
It is observed that the ``would be''
zero modes separate from the non-chiral modes with
higher level of improvement. In addition, as the level
of improvement increases, the continuum 4-fold degeneracy
emerges where the scattered eigenmodes begin to form quartets
(remember that only half of the spectrum is computed
so the data points form doublets). Note also that at this
coupling the Asq operator still retains large lattice
artifacts so the ``would be'' zero modes cannot be
identified. Further improvement is required.

\begin{figure}
\begin{minipage}[b]{8.5pc}
\psfrag{gname}[rb][rb]{$\mathrm{UFat7}\times\mathrm{Asq}$}
\psfrag{beta}[rb][rb]{$\beta=6.0$}
\psfrag{xaxis}[ct][ct]{\tiny $\lambda$}
\psfrag{yaxis}[cb][cb]{\tiny $|\chi|$}
\includegraphics[width=5.9pc,angle=-90]{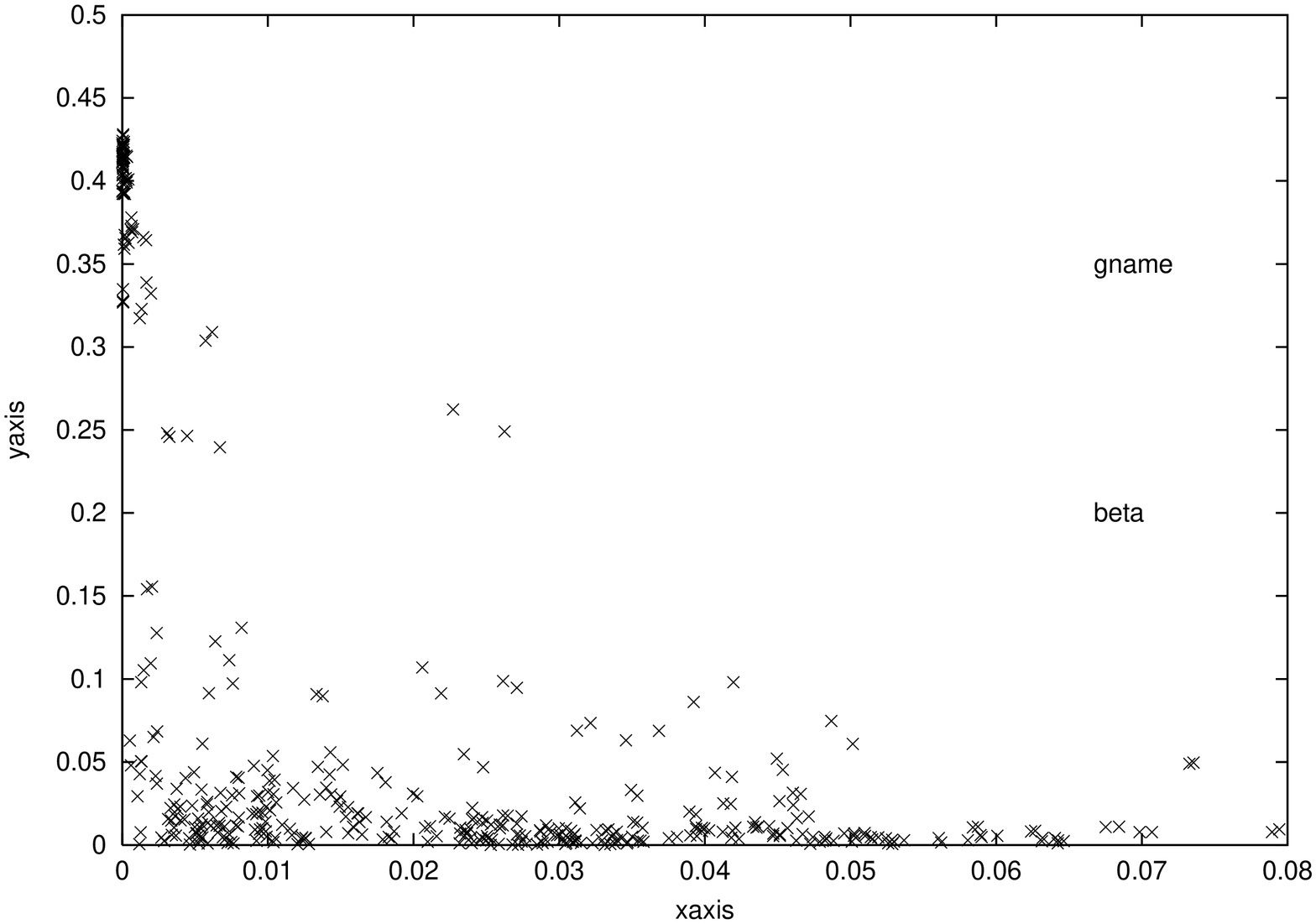}
\end{minipage}%
\begin{minipage}[b]{8.5pc}
\psfrag{gname}[rb][rb]{$\mathrm{UFat7}\times\mathrm{Asq}$}
\psfrag{beta}[rb][rb]{$\beta=6.2$}
\psfrag{xaxis}[ct][ct]{\tiny $\lambda$}
\psfrag{yaxis}[cb][cb]{\tiny $|\chi|$}
\includegraphics[width=5.9pc,angle=-90]{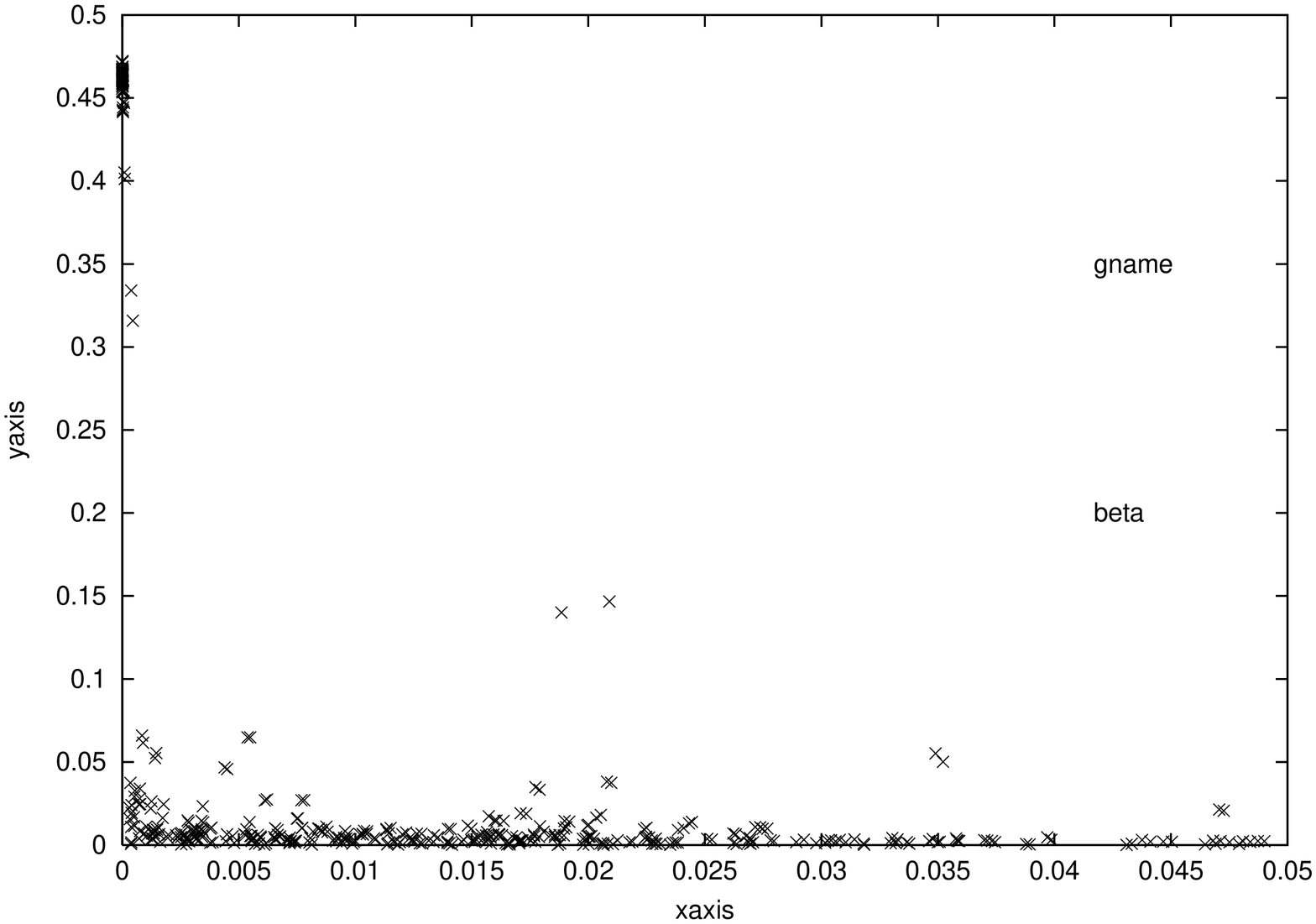}
\end{minipage}
\caption{Spectra at $\beta=6.0$ and $6.2$.}
\label{betaspectrum}
\end{figure}

\begin{figure}
\begin{minipage}[b]{8.5pc}
\psfrag{xaxis}[ct][ct]{\tiny Configuration}
\psfrag{yaxis}[cb][cb]{\tiny Ratio}
\psfrag{fatasq}[rb][rb]{\tiny $\mathrm{UFat7}\times\mathrm{Asq}$}
\psfrag{hyp}[rb][rb]{\tiny $\mathrm{HYP}$}
\psfrag{overlap}[rb][rb]{\tiny overlap}
\psfrag{beta}[lb][lb]{\small $\beta=5.85$}
\includegraphics[width=5.9pc,angle=-90]{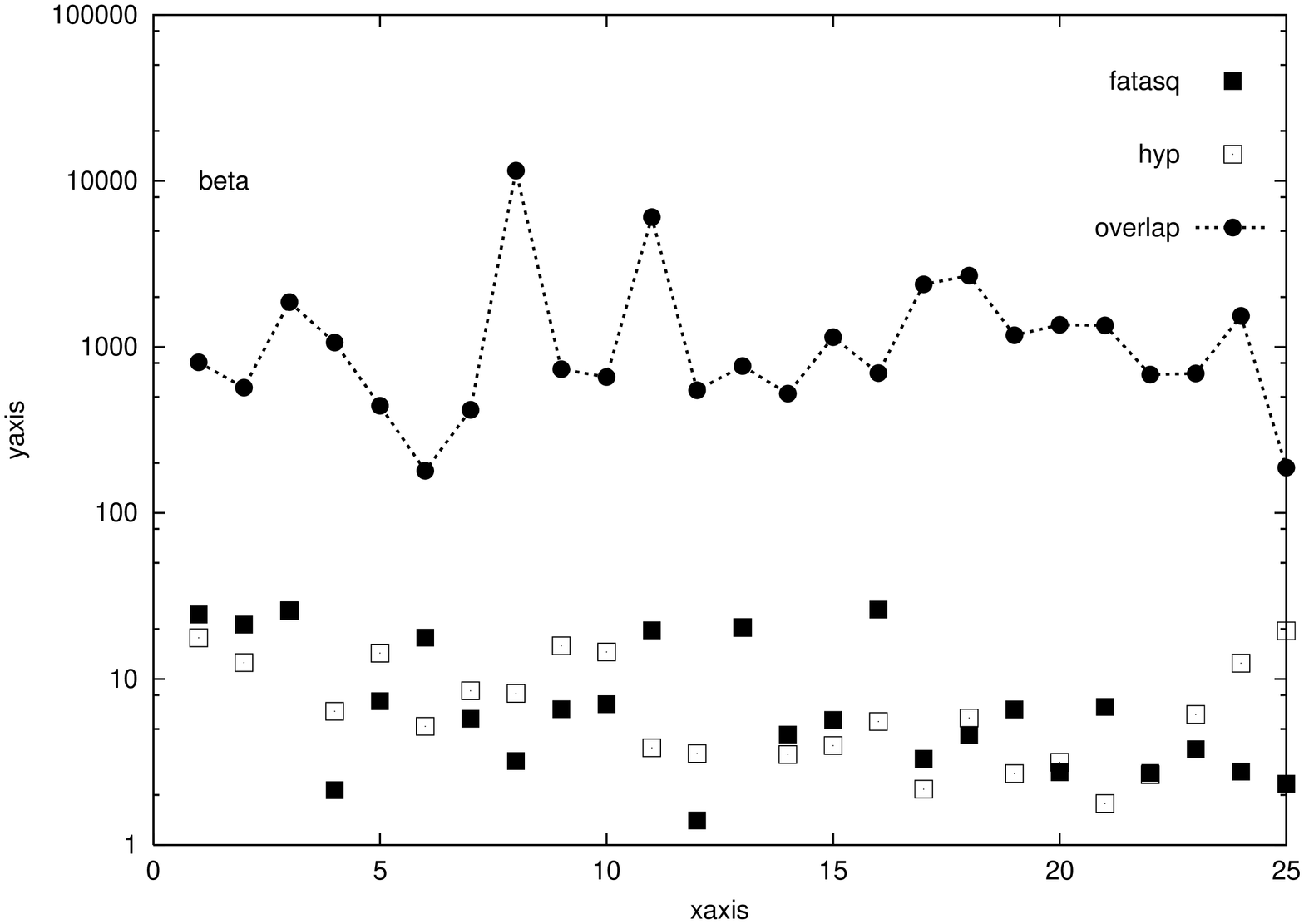}
\end{minipage}%
\begin{minipage}[b]{8.5pc}
\psfrag{xaxis}[ct][ct]{\tiny Configuration}
\psfrag{yaxis}[cb][cb]{\tiny Ratio}
\psfrag{fatasq}[rb][rb]{\tiny $\mathrm{UFat7}\times\mathrm{Asq}$}
\psfrag{hyp}[rb][rb]{\tiny $\mathrm{HYP}$}
\psfrag{beta}[lb][lb]{\small $\beta=6.2$}
\includegraphics[width=5.9pc,angle=-90]{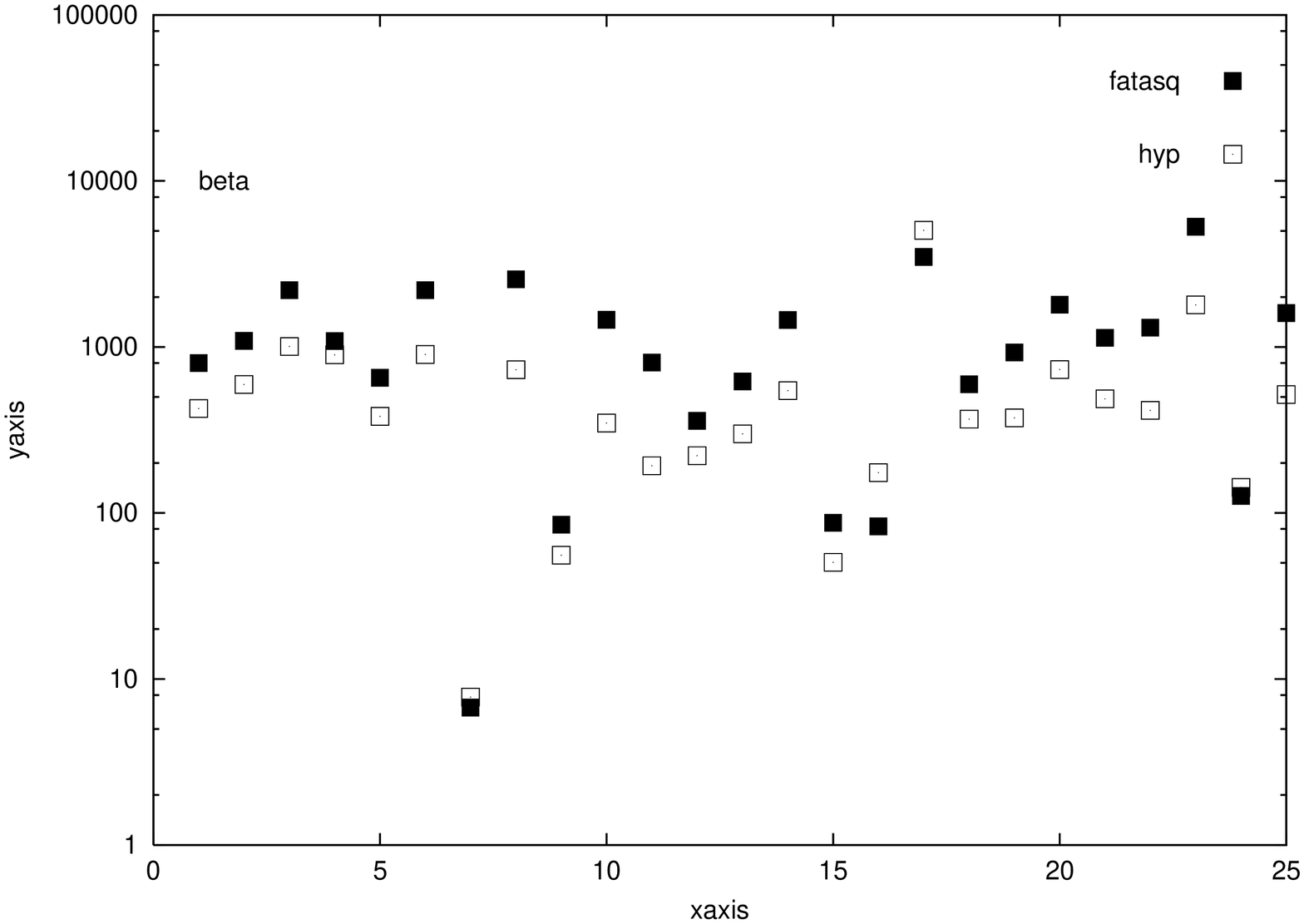}
\end{minipage}
\caption{Ratio between the eigenvalues of the smallest
non-chiral modes and the largest ``zero mode''.}
\label{ratioplot}
\end{figure}

We next study the dependence of the eigenvalue spectrum
on $\beta$. The spectra of the $\mathrm{UFat7}\times\mathrm{Asq}$
operator at $\beta=6.0$ and $6.2$ are given in
figure \ref{betaspectrum}. Again, the ``would be'' zero
modes are well separated from the non-chiral modes
and the continuum 4-fold degeneracy is better realized
as one approaches the continuum limit.
To quantify the separation between the ``zero modes''
and the non-chiral modes, the ratio of eigenvalues
between the smallest non-chiral mode and the largest ``zero
mode'' is plotted in figure \ref{ratioplot}. The ratios increase
from one order of magnitude at $\beta=5.85$ to three orders
of magnitude at $\beta=6.2$ for the improved staggered
operators. Theoretically, the ratio is infinite for the
overlap operator because exact zero modes exist on the lattice
for overlap fermions. It is finite here solely because of computational
precision. Results here also show that improvement using
fat-links or hyper-cubic blocking are equally efficient.

\begin{figure}
\begin{minipage}[b]{5.8pc}
\psfrag{gname}[lb][lb]{\small (a)}
\psfrag{xaxis}[ct][ct]{\tiny $Q$}
\psfrag{yaxis}[cb][cb]{\tiny Percentage}
\includegraphics[width=3.9pc,angle=-90]{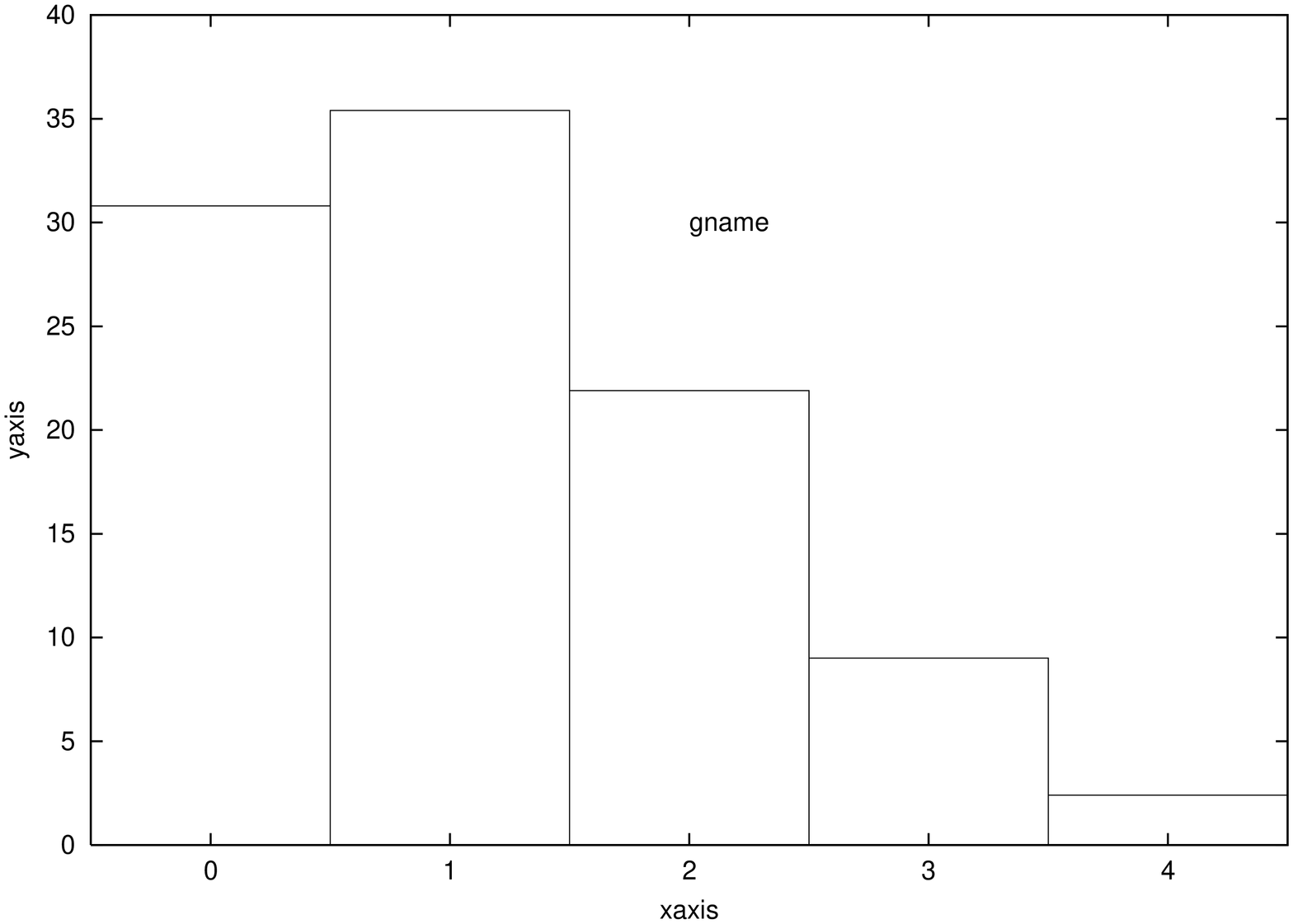}
\end{minipage}%
\begin{minipage}[b]{5.8pc}
\psfrag{gname}[lb][lb]{\small (b)}
\psfrag{xaxis}[ct][ct]{\tiny $Q$}
\psfrag{yaxis}[cb][cb]{\tiny Percentage}
\includegraphics[width=3.9pc,angle=-90]{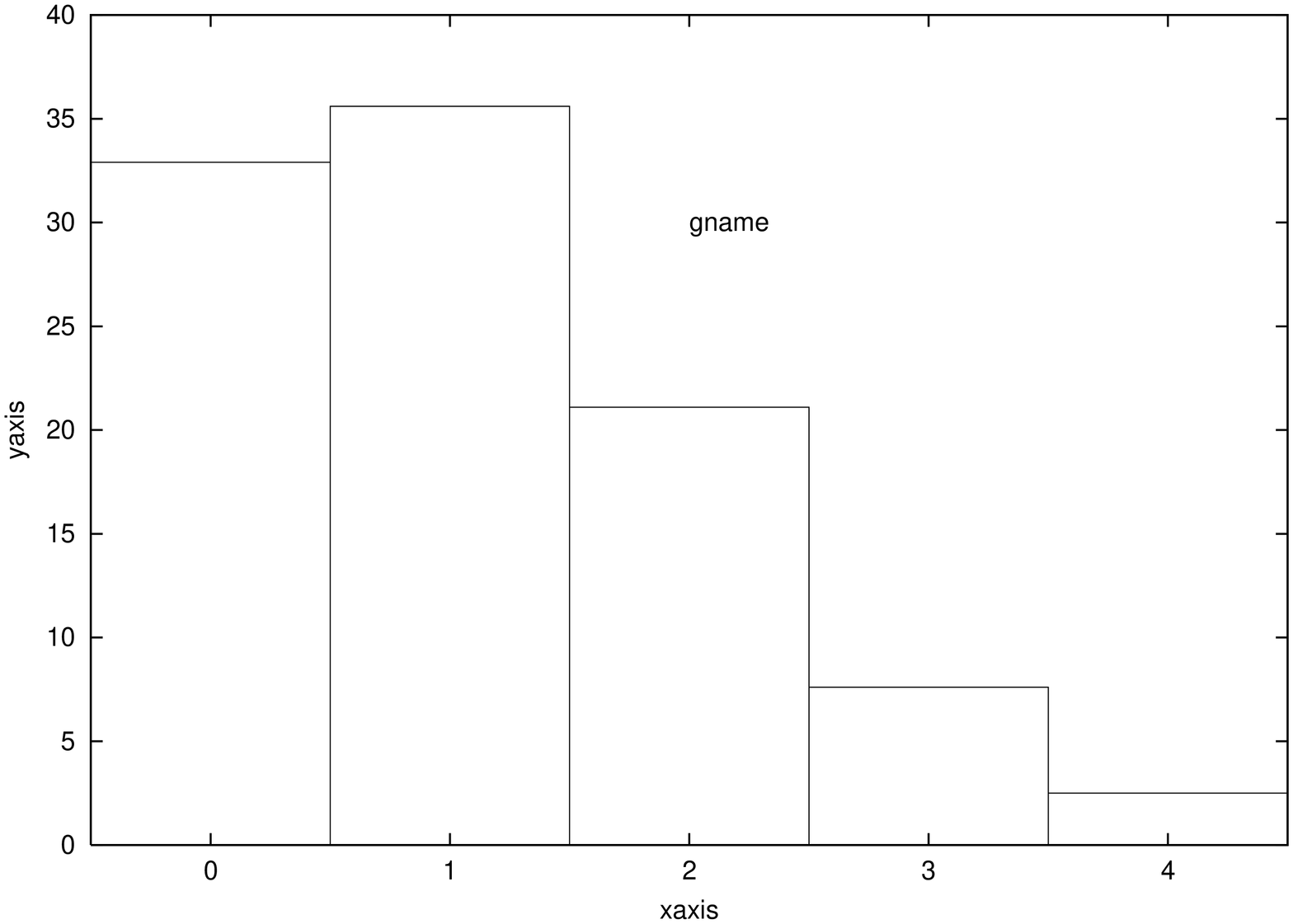}
\end{minipage}%
\begin{minipage}[b]{5.8pc}
\psfrag{gname}[lb][lb]{\small (c)}
\psfrag{xaxis}[ct][ct]{\tiny $Q$}
\psfrag{yaxis}[cb][cb]{\tiny Percentage}
\includegraphics[width=3.9pc,angle=-90]{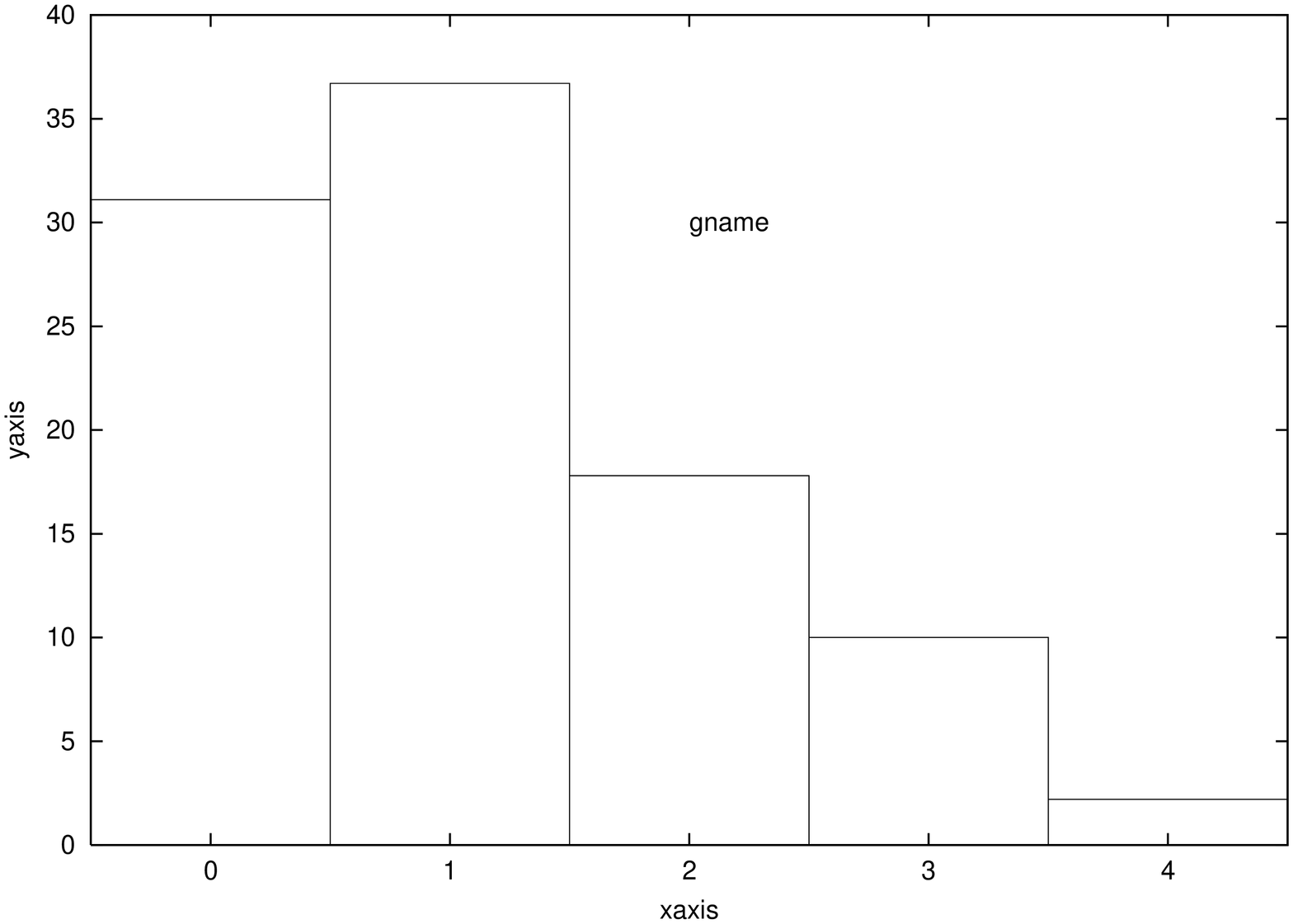}
\end{minipage}%
\caption{Normalized topological charge distributions
(a) $\mathrm{UFat7}\times\mathrm{Asq}$
(b) $\mathrm{HYP}$ (c) overlap.}
\label{qdist}
\end{figure}

Because of lattice artifacts, the topological charge indices
obtained by using different operators do not agree on a
configuration by configuration basis. In the present case we
find that the charge indices determined by using different
operators agree at about 60-70\%, compared to 28\% if the values
were completely random. It is then important to check whether the
charge distributions are also different because physical
observables, e.g., the susceptibility, are related to the
average of the topological charge. The distributions obtained
by three different operators are given in figure \ref{qdist}.
It can be observed that there is no significant difference among
the results. Consequently, one would expect topological
quantities obtained by these operators to agree.

\begin{figure}
\begin{minipage}[b]{8.5pc}
\psfrag{gname}[rb][rb]{\small $\mathrm{UFat7}\times\mathrm{Asq}$}
\psfrag{beta}[rb][rb]{\small $\beta=5.85$}
\psfrag{v0}[cb][cb]{\tiny $Q=0$}
\psfrag{v1}[cb][cb]{\tiny $Q=1$}
\psfrag{v2}[cb][cb]{\tiny $Q=2$}
\psfrag{xaxis}[ct][ct]{\tiny $z$}
\psfrag{yaxis}[cb][cb]{\tiny Probability}
\includegraphics[width=5.9pc,angle=-90]{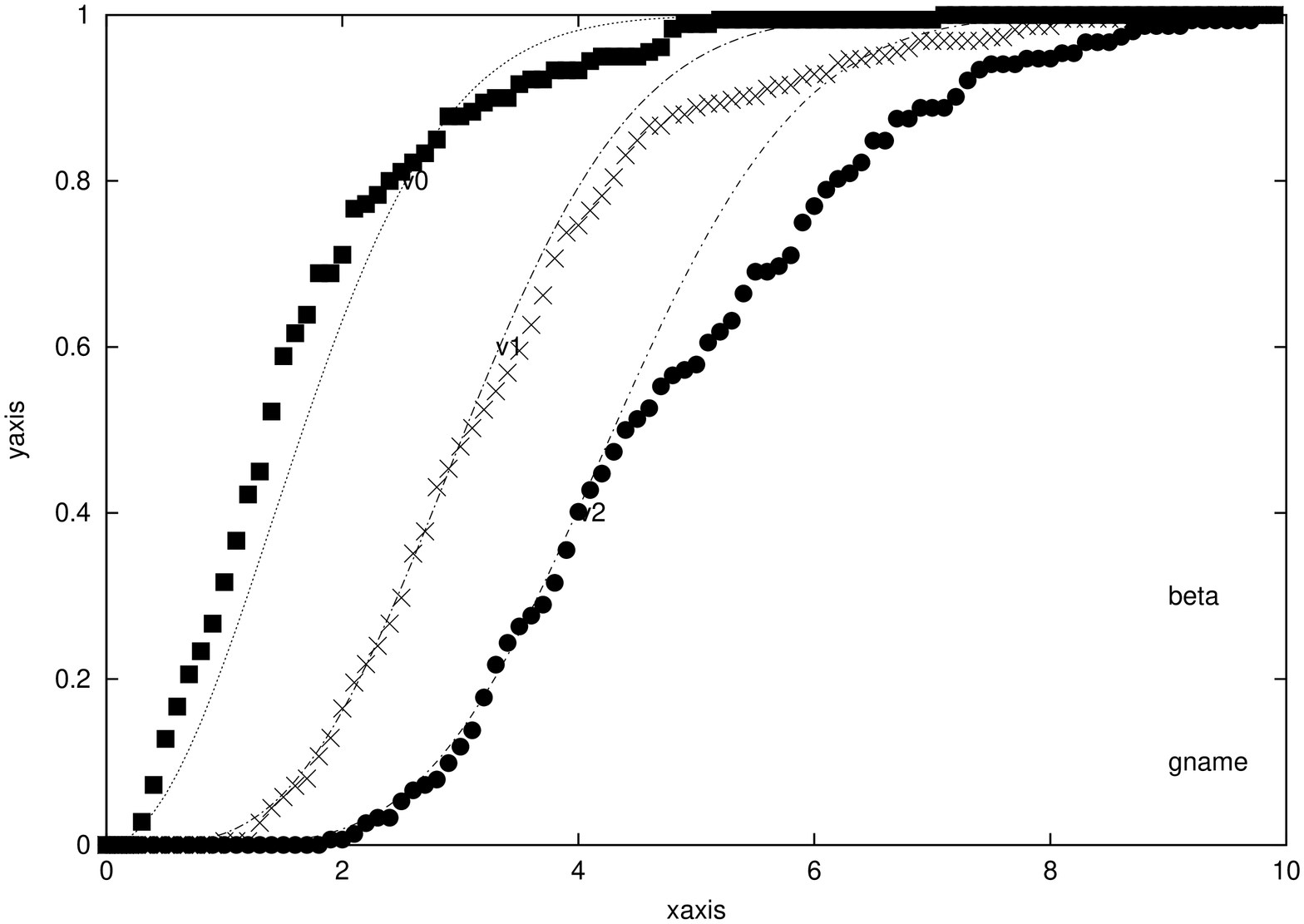}
\end{minipage}%
\begin{minipage}[b]{8.5pc}
\psfrag{gname}[rb][rb]{\small overlap}
\psfrag{beta}[rb][rb]{\small $\beta=5.85$}
\psfrag{v0}[cb][cb]{\tiny $Q=0$}
\psfrag{v1}[cb][cb]{\tiny $Q=1$}
\psfrag{v2}[cb][cb]{\tiny $Q=2$}
\psfrag{xaxis}[ct][ct]{\tiny $z$}
\psfrag{yaxis}[cb][cb]{\tiny Probability}
\includegraphics[width=5.9pc,angle=-90]{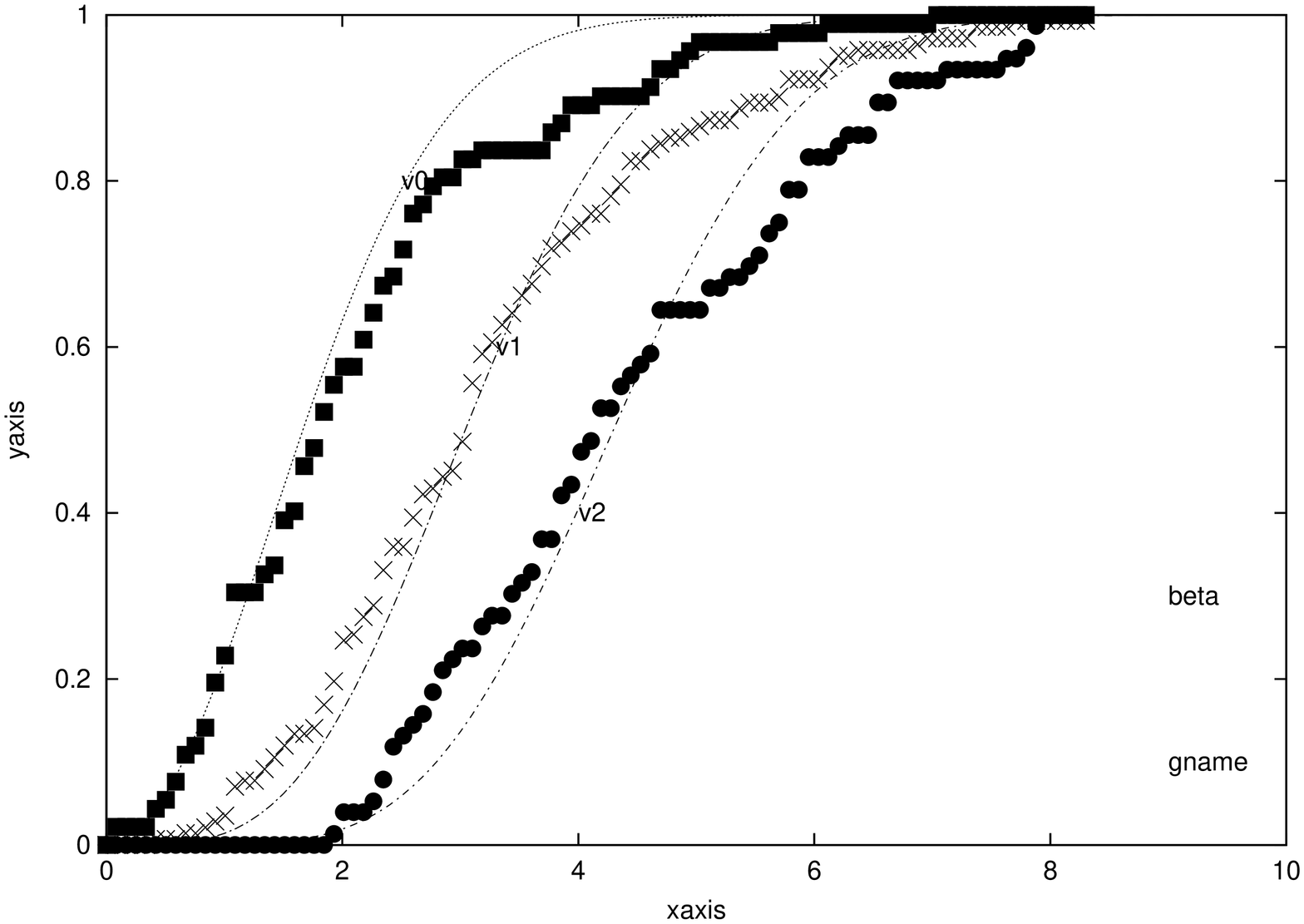}
\end{minipage}%
\caption{The cumulative distributions of the smallest
non-chiral mode in different topological sectors. Solid lines are
predictions of RMT. The eigenvalues are rescaled by
the infinite-volume chiral condensate $\Sigma$, 
$z=\lambda V \Sigma$, where $\Sigma$ is a fitting
parameter here.}
\label{rmtgraph}
\end{figure}

Finally distribution of the remaining non-chiral modes
is compared with the predictions of RMT. In particular,
we focus on the cumulative distribution of the smallest
non-chiral modes \cite{Damgaard2000}. The results are shown in
figure \ref{rmtgraph}. Agreement with the predictions of RMT is
evident and is comparable to that obtained with overlap fermions\cite{bieten}.

In summary, we have shown that the eigenvalue spectrum of the
staggered Dirac operator is very sensitive to lattice
artifacts. Better continuum properties are obtained by using
improved staggered operators.
The topological charge distributions calculated
using different operators
show no significant difference. Finally, distribution
of the smallest non-chiral mode is compared with the predictions
of RMT and satisfactory agreement is obtained.

Two related works with similar conclusions
were also reported at this conference \cite{Follana2004}.

We are very grateful to J. B. Zhang for providing the
eigenvalue solver for the overlap operator and to H. Trottier
for many useful discussions. The computations were
performed using WestGrid facilities. 
This work is supported in part by the Natural Sciences and
Engineering Research Council of Canada.

\end{document}